\documentclass[aps,prb,twocolumn,10pt,amsmath,amssymb,reprint,floatfix,superscriptaddress]{revtex4-1}
\usepackage{amsmath,amssymb}
\usepackage{graphicx}

\newcommand{\quant}[2]{$#1\,\text{#2}$}

\begin{document}

\title{Role of vibrational entropy in the stabilization of the high-temperature phases of iron}

\author{J\"urgen Neuhaus}
\author{Michael Leitner}
\affiliation{Chair of Functional Materials, Physics Department, Technische Universit\"at M\"unchen, James-Franck-Str.\ 1, 85748 Garching, Germany}
\affiliation{Heinz Maier-Leibnitz Zentrum (MLZ), Technische Universit\"at M\"unchen, Lichtenbergstr.\ 1, 85748 Garching, Germany}
\author{Karl Nicolaus}
\affiliation{Chair of Functional Materials, Physics Department, Technische Universit\"at M\"unchen, James-Franck-Str.\ 1, 85748 Garching, Germany}
\author{Winfried Petry}
\affiliation{Chair of Functional Materials, Physics Department, Technische Universit\"at M\"unchen, James-Franck-Str.\ 1, 85748 Garching, Germany}
\affiliation{Heinz Maier-Leibnitz Zentrum (MLZ), Technische Universit\"at M\"unchen, Lichtenbergstr.\ 1, 85748 Garching, Germany}
\author{Bernard Hennion}
\affiliation{Laboratoire L\'eon Brillouin, CEA Saclay, 91191 Gif-sur-Yvette C\'edex, France}
\author{Arno Hiess}
\affiliation{Institut Laue-Langevin, 38042 Grenoble, France, now at European Spallation Source AB, 22100 Lund, Sweden}

\begin{abstract}
The phonon dispersions of the bcc and fcc phases of pure iron ($\alpha$-Fe, $\gamma$-Fe and $\delta$-Fe) at ambient pressure were investigated close to the respective phase transition temperatures. In the open bcc structure the transverse phonons along T$_1[\xi\xi 0]$ and T$_1[\xi\xi2\xi]$ are of particularly low energy. The eigenvectors of these phonons correspond to displacements needed for the transformation to the fcc $\gamma$-phase. Especially these phonons, but also all other phonons soften considerably with increasing temperature. Comparing thermodynamic properties of the fcc and the two bcc phases it is shown that the high temperature bcc phase is stabilized predominantly by vibrational entropy, whereas for the stabilization of the fcc phase electronic entropy provides an equal contribution. 
\end{abstract}


\maketitle

\section{Introduction}
The majority of metals crystallize from the melt in the open body-centered cubic structure, and the largest part of those transforms martensitically to a close-packed structure at lower temperatures. While the latter fact has been understood since a long time as the optimal solution to the electrostatic attraction between valence electrons and ionic cores subject to closed-shell repulsion \cite{fuchsprocroysoca1935}, proposals to motivate the preference for the open structure at high temperatures have been controversial. The classical explanations due to Zener \cite{zenerpr1947} and Friedel \cite{friedeljphyslett1974} identify the gain in vibrational entropy due to specific low-energy phonons to be responsible, a view which is shared by the majority of later numerical calculations \cite{yeprl1987,rubiniprb1993,osetskyprb1998,daiscience2003,souvatzkisprl2008} (but see Ref.~\onlinecite{willaimeprl1989} for a conflicting result). Direct experimental studies, where available, confirm the dominant contribution of vibrational entropy for the thermodynamic stability of bcc structures at high temperatures, e.g., for the transition metals of group 3 and 4 \cite{petryprb1991,heimingprb1991,guethoffprb1993,petryjphysiv1995}, and the general necessity of considering the role of vibrations in the thermodynamics of materials, specifically at high temperatures, is recognized \cite{fultzprogmatersci2010}.

The polymorphism of iron, a system with strong magnetic interactions, is of particular interest in this context. Pure iron solidifies at \quant{1811}{K} in the bcc $\delta$-phase and undergoes a first transition to fcc $\gamma$-Fe at \quant{1667}{K}. Very unusual, with decreasing temperature it transforms back to the bcc $\alpha$-phase at \quant{1185}{K}. Within the $\alpha$-phase a magnetic transition occurs at \quant{1043}{K}, below of which $\alpha$-Fe is ferromagnetic \cite{chenjphaseequilbria2001}.

The occurrence of the bcc $\alpha$-phase is understood in the framework of band magnetism \cite{hasegawaprl1983} as being due to ferromagnetic contributions to the total energy, which can be reproduced by density-functional calculations in the generalized-gradient approximation \cite{singhprb1991,herperprb1999}. In contrast to the proposal of the 2$\gamma$-state model \cite{kaufmanactamet1963}, the dominant view nowadays is that with the loss of magnetic correlations at higher temperatures a single paramagnetic fcc $\gamma$-phase results, as it corresponds to the non-magnetic structure of lowest total energy \cite{singhprb1991,leonovprl2011}. The small region of paramagnetic $\alpha$-Fe is thought to be due to the persistence of local moments even above $T_\text{C}$ \cite{leonovprl2011,anisimovprb2012}.

More controversial are explanations for the $\gamma\to\delta$ transition, i.e., the question why Fe adopts again bcc in its high temperature phase. Published records of theoretical calculations focussing either purely on the electronic \cite{hasegawaprl1983} or vibrational \cite{osetskyprb1998,sandovalprb2009} contribution to the entropy generally find that the effect considered in the respective studies suffices for explaining the observed behaviour, while semi-empirical fits to experimental data \cite[e.g.,][]{kaufmanactamet1963,grimvallphysscr1976} favor electronic reasons (but see Ref.~\onlinecite{grimvallphysscr1976} for an overview of the widely differing assumptions in such approaches). Part of the interest in the $\delta$-phase is due to the fact that a paramagnetic bcc structure with reduced lattice constant is also proposed for the earth inner core, stabilized by vibrational entropy \cite{yooprl1993,vovcadlonature2003,belonoshkonature2003,luopnas2010}.

By measuring the phonon dispersion of $\delta$-Fe for the first time together with temperature-dependent dispersions for $\alpha$- as well as $\gamma$-Fe we are able to determine the vibrational entropy of the distinct ambient-pressure iron phases purely from experiment and to evaluate the role of vibrational entropy in stabilizing the high-temperature phases.

\begin{figure*}[t]
  \begin{minipage}[t]{10.61cm}
    \vspace{0pt}
    \includegraphics{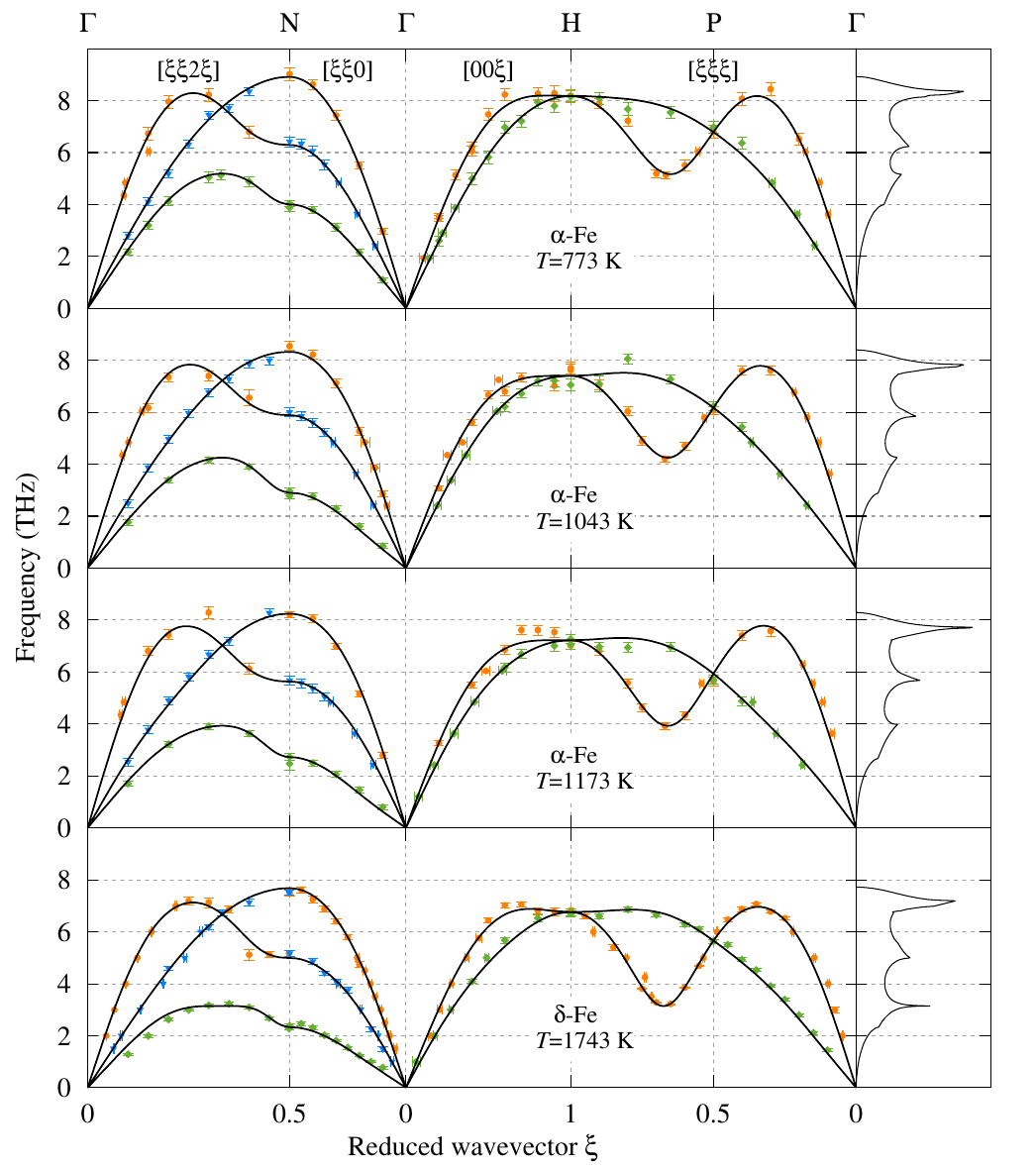}
  \end{minipage}
  \begin{minipage}[t]{7.09cm}
    \vspace{0pt}
    \includegraphics{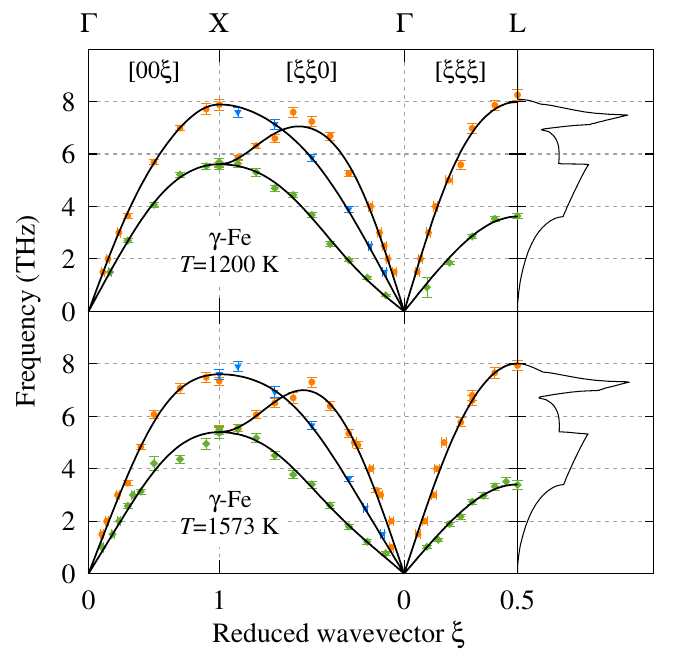}
    \caption{Phonon dispersions of Fe. The solid lines correspond to the expected values of the Born-von K\'arm\'an parameters' a posteriori distribution, from which the densities of states have been calculated. Circles (orange) represent measured longitudinal modes, diamonds (green) transverse modes. Triangles (blue) correspond to the T$_2[\xi\xi 0]$ and T$_2[\xi\xi 2\xi]$ branches, where appropriate.}\label{disp}
  \end{minipage}
\end{figure*}

\section{Experimental details}
Several large single crystals of the $\delta$-phase with a typical size of 40 to \quant{50}{mm} in length and \quant{10}{mm} in diameter were grown from high purity (4N) Fe rods by the zone melting technique using our combined single crystal growth and measuring furnace \cite{flottmannnima1987}. After the {\em in-situ} growth on the three-axis spectrometer they were kept continuously above the transition temperature $T_{\gamma\leftrightarrow\delta}$. To suppress evaporation of the sample over the course of the measurement (the evaporation rate under vacuum at \quant{1743}{K} was estimated to \quant{20}{g/h}) a high purity Argon atmosphere of \quant{700}{mbar} was used. The temperature could be stabilized within \quant{\pm 5}{K} with a gradient along the single crystalline part of the sample of about \quant{15}{K}. The absolute temperature was calibrated by the known transition temperatures $T_{\alpha\leftrightarrow\gamma}$ and $T_{\gamma\leftrightarrow\delta}$.

High purity single crystals of the $\alpha$-phase with \quant{5.5}{mm} in diameter and variable length were grown by recrystallization at the Max-Planck-Institut f\"ur Metallforschung, Stuttgart. A standard resistance furnace has been used to heat the crystals under vacuum with an accuracy of \quant{\pm 8}{K}.  Heating these $\alpha$-Fe single crystals into the $\gamma$-phase transformed them to a nearly perfect {\em powder} (polycrystalline) sample. However, cycling approximately ten times through the $\alpha$-$\gamma$ transition finally led to the growth of a \quant{3}{cm$^3$} $\gamma$-single crystal by recrystallization.

The measurements in the $\delta$-phase were performed at the three-axis spectrometer 1T at the LLB, Saclay, in the $\alpha$-phase at spectrometer E7 at the HMI, Berlin and those in the $\gamma$-phase at spectrometer IN3 at the ILL, Grenoble. For all measurements a pyrolytic graphite monochromator and analyzer were used in constant final wave-vector mode.

\section{Phonon dispersions}
The phonon dispersions of bcc iron were measured at \quant{773}{K}, \quant{1043}{K}, \quant{1173}{K} (preliminarily published in Ref.~\onlinecite{neuhausphysb1997}) and \quant{1743}{K}. In the $\gamma$-phase experiments were done at \quant{1200}{K} and \quant{1573}{K}. All measurements cover the main symmetry directions and, for the bcc phases, additionally the $[\xi\xi2\xi]$ direction. The obtained phonon frequencies are presented in figure \ref{disp} (see the supplemental material \cite{supp} for the data). Within the bcc-phases a softening of the entire phonon dispersion is observed when passing from room temperature (cp.~Refs.~\onlinecite{brockhousesolidstatecomm1967,bergsmaphysletta1967,minkiewiczpr1967,vandijkneutron1968,klotzprl2000}) to $T_{\alpha\leftrightarrow\gamma}$. Most pronounced, however, is the decrease of the transverse branches T$_1[\xi\xi2\xi]$ and T$_1[\xi\xi0]$, reducing to a value of 53\% at the zone boundary in the $\delta$-phase compared to room temperature. This softening has a nonlinear temperature dependence, particularly around the ferromagnetic transition as observed earlier \cite{vallerathesis1977,*vallerajphyscoll1981,satijaprb1985}. Interestingly, recent finite-temperature ab-initio calculations show that a variety of independent phenomena can give rise to this effect: dynamical mean-field theory \cite{leonovprb2012} (treating electronic excitations), density-functional theory under disordered local moment paramagnetism \cite{koermannprb2012} as well as self-consistent lattice dynamics \cite{luopnas2010} (treating anharmonic effects) agree that the softening with temperature is strongest for this branch. Our data also display an increase of the linewidth of these phonons with increasing temperature in the order of \quant{0.1}{THz}, however the damping of transverse phonons in the $\alpha$-phase is considerably smaller than in the $\delta$-phase. 

\begin{table}[t]
  \setlength{\tabcolsep}{1pt}
\caption{Force constants in N/m estimated by Bayesian inference from the phonon dispersion of Fe at various temperatures in the bcc phases with a Born-von K\'arm\'an model taking into account interactions up to the fifth neighbour shell.}\label{bvkbcc}
\renewcommand{\arraystretch}{.8}
\begin{tabular}{l@{\hskip 1em}rl@{\hskip 1em}rl@{\hskip 1em}rl@{\hskip 1em}rl}
  \hline\hline
  & \rlap{\quant{773}{K}}\hphantom{aaa}& & \rlap{\quant{1043}{K}}\hphantom{aaaa}& & \rlap{\quant{1173}{K}}\hphantom{aaaa}& & \rlap{\quant{1743}{K}}\hphantom{aaaa}&\\
  \hline
$\Phi^{[111]}_\text{L}$&44.53&(49)&41.99&(37)&41.23&(36)&35.95&(21)\\
$\Phi^{[111]}_\text{T}$&1.10&(68)&-1.99&(50)&-3.35&(49)&-2.76&(24)\\[5pt]
$\Phi^{[200]}_\text{L}$&11.44&(90)&7.50&(62)&7.51&(61)&9.34&(37)\\
$\Phi^{[200]}_\text{T}$&0.18&(46)&0.65&(34)&-0.14&(34)&-0.98&(19)\\[5pt]
$\Phi^{[220]}_\text{L}$&2.91&(42)&2.72&(33)&3.20&(31)&0.85&(16)\\
$\Phi^{[220]}_{\text{T}[1\overline{1}0]}$&-0.27&(31)&-0.04&(25)&-0.48&(25)&1.07&(10)\\
$\Phi^{[220]}_{\text{T}[001]}$&-0.60&(45)&-0.16&(34)&0.27&(35)&-0.87&(24)\\[5pt]
$\Phi^{[311]}_{xx}$&-0.15&(22)&0.12&(16)&0.19&(17)&-0.18&(10)\\
$\Phi^{[311]}_{yy}$&-0.07&(15)&-0.13&(12)&0.12&(11)&0.25&(7)\\
$\Phi^{[311]}_{yz}$&-0.16&(21)&-0.24&(16)&-0.33&(16)&-0.22&(5)\\
$\Phi^{[311]}_{xy}$&-0.09&(15)&0.24&(12)&0.34&(11)&0.22&(4)\\[5pt]
$\Phi^{[222]}_\text{L}$&0.50&(26)&0.50&(22)&0.47&(21)&1.56&(11)\\
$\Phi^{[222]}_\text{T}$&-0.34&(47)&0.13&(36)&0.18&(34)&-0.40&(18)\\
  \hline
\end{tabular}
\end{table}

Concerning phonon anomalies, i.e., low frequencies and strong damping along T$_1[\xi\xi 0]$ and T$_1[\xi\xi2\xi]$, the dispersion of $\delta$-Fe resembles to the high temperature phases of group 3 (including some lanthanides) and 4 metals \cite{petryprb1991,heimingprb1991,guethoffprb1993,trampenauprb1991,petryprb1993,nicolauseurphysjb2001}. These anomalies are indicative for latent instabilities towards a transition to a close-packed structure \cite{petryjphysiv1995}. As in the case of La and Ce, Fe transforms to the fcc structure, in contrast to the other group 3 and 4 metals, which transform to hcp. The resulting stacking sequence of close-packed planes necessitates long-wavelength shears, with the corresponding shear constant $C'=1/2(C_{11}-C_{12})$ given by the initial slope of the T$_1[\xi\xi0]$ phonon branch. Indeed, Table \ref{therm} shows that the cubic anisotropy parameter $A=C_{44}/C'$ (i.e., the squared ratio of the small-$q$ slopes of the $[\xi\xi0]$ acoustic branches), which has values below 6 in the elements transforming to hcp \cite{petryprb1991,heimingprb1991,petryprb1993}, reaches up to around 10 for Fe, La \cite{petryjphysiv1995} and Ce \cite{nicolauseurphysjb2001}, analogous to the criteria put forward in Ref.~\onlinecite{willsprl1992}. For comparison, Cr \cite{trampenauprb1993} and Nb \cite{guethoffjphyscondmat1994}, which display the bcc structure over the whole range, have an anisotropy parameter on the order of one. The most remarkable point concerning the $\gamma$-phase dispersions is the positive curvature of the T$_1[\xi\xi0]$ branch, which reproduces the findings of Zarestky and Stassis \cite{zarestkyprb1987}.

\section{Data modelling and Born-von K\'arm\'an parameters}
For deducing further quantitative information we describe the measured phonon dispersions by a Born-von K\'arm\'an model (corresponding to the quasi-harmonic assumption). In order to obtain methodically rigorous uncertainties of the estimated quantities, we followed Bayesian inference and generated samples of the Born-von K\'arm\'an force constants including interactions up to the fifth nearest neighbor shell for each measured phonon dispersion. We computed the likelihood directly from the experimentally estimated errors. As the information contained in measurements of the high-symmetry directions alone is limited (see for example the pertinent discussion in Ref.~\onlinecite{svenssonpr1967}), we used a prior distribution that penalizes high values of the force constants for far shells and non-central forces, as dictated by physical understanding (see supplemental material \cite{supp}). 

Phonon dispersions corresponding to the resulting expected values of the force constants are shown in figure \ref{disp} as solid lines, while an illustration of the uncertainties is given in the supplemental material \cite{supp}. The Born-von K\'arm\'an force constants (mean and standard deviation) are summarized in Tabs.~\ref{bvkbcc} and \ref{bvkfcc} for the bcc and fcc measurements, respectively. For those sites that are along high-symmetry directions relative to the central atom, we parametrized the model directly in terms of longitudinal and transversal force constants, that is in terms of the eigenvalues of the Jacobi matrices of the forces, while for lower-symmetry shells we give the independent entries with respect to the Cartesian basis of the Jacobi matrices. The parameters for the respective shells' other sites follow by symmetry.

\begin{table}[t]
  \setlength{\tabcolsep}{1pt}
\caption{Force constants in N/m estimated by Bayesian inference from the phonon dispersion of Fe at various temperatures in the fcc phases with a Born-von K\'arm\'an model taking into account interactions up to the fifth neighbour shell.}\label{bvkfcc}
\begin{tabular}{l@{\hskip 1em}rl@{\hskip 1em}rl}
  \hline\hline
  & \rlap{\quant{1200}{K}}\hphantom{aaa}& & \rlap{\quant{1573}{K}}\hphantom{aaaa}&\\
  \hline
$\Phi^{[110]}_\text{L}$&30.16&(37)&29.84&(41)\\
$\Phi^{[110]}_{\text{T}[1\overline{1}0]}$&-2.27&(48)&-2.16&(48)\\
$\Phi^{[110]}_{\text{T}[001]}$&0.18&(79)&-1.26&(83)\\[5pt]
$\Phi^{[200]}_\text{L}$&-1.90&(74)&-0.21&(74)\\
$\Phi^{[200]}_\text{T}$&0.34&(35)&0.15&(35)\\[5pt]
$\Phi^{[211]}_{xx}$&0.18&(28)&0.61&(29)\\
$\Phi^{[211]}_{yy}$&-0.01&(19)&-0.04&(20)\\
$\Phi^{[211]}_{yz}$&0.24&(17)&0.27&(17)\\
$\Phi^{[211]}_{xy}$&0.41&(10)&0.35&(12)\\[5pt]
$\Phi^{[220]}_\text{L}$&0.85&(30)&0.71&(30)\\
$\Phi^{[220]}_{\text{T}[1\overline{1}0]}$&0.14&(32)&0.25&(32)\\
$\Phi^{[220]}_{\text{T}[001]}$&-0.01&(43)&-0.21&(43)\\[5pt]
$\Phi^{[310]}_{xx}$&0.27&(13)&-0.22&(14)\\
$\Phi^{[310]}_{yy}$&0.05&(22)&-0.32&(23)\\
$\Phi^{[310]}_{zz}$&-0.19&(24)&0.07&(24)\\
$\Phi^{[310]}_{xy}$&0.14&(20)&-0.30&(21)\\
  \hline
\end{tabular}
\end{table}

The behaviour of the determined BvK-parameters is quite plausible: As expected, the dominant interactions are short-range and of longitudinal nature, while most of the interactions over longer ranges are individually not significantly different from zero (collectively, they are significant, however; setting all of them to zero would give noticeably worse fits). Apart from the softening of the nearest-neighbour longitudinal interaction with temperature, the most striking effect is the behaviour of $\Phi^{[111]}_\text{T}$ in the bcc phases: its becoming negative with increasing temperature is the main reason for the softening of the T$_1[\xi\xi0]$ branch. As this happens around the Curie temperature, it is most probably of magnetic origin and pinpoints the instability of the body-centered lattice with loss of magnetism. Note that the $\delta$-phase's comparatively larger long-range BvK-parameters should not be over-interpreted, as they are due to the better statistics of this one measurement (also reflected in the smaller uncertainties).

The generated samples of parameter sets can be used to directly compute thermodynamic quantities. By virtue of this approach, we can quote well-defined estimated errors for the computed quantities that follow directly from the errors of the experimental data points, subject only to the assumption of the Born-von K\'arm\'an model and the chosen prior distribution. Figure \ref{disp} demonstrates that we do not over-regularize our data, therefore we are confident that our quoted estimated errors are conservative. 

The phonon densities of states for each of the measured temperatures as calculated from the force constants by the tetrahedron method \cite{lehmannpss1970,*lehmannpssb1972,jepsonsolidstatecomm1971} are given to the right of the phonon dispersions in figure \ref{disp}. They show that, according to the model, the softening is not limited to the measured high-symmetry directions, but spans the whole reciprocal space. Note that the spike appearing slightly above \quant{3}{THz} in the $\delta$-phase is due to the flattening of the T$_1[\xi\xi2\xi]$ branch.

\section{Thermodynamic quantities}
\begin{figure}
  \includegraphics{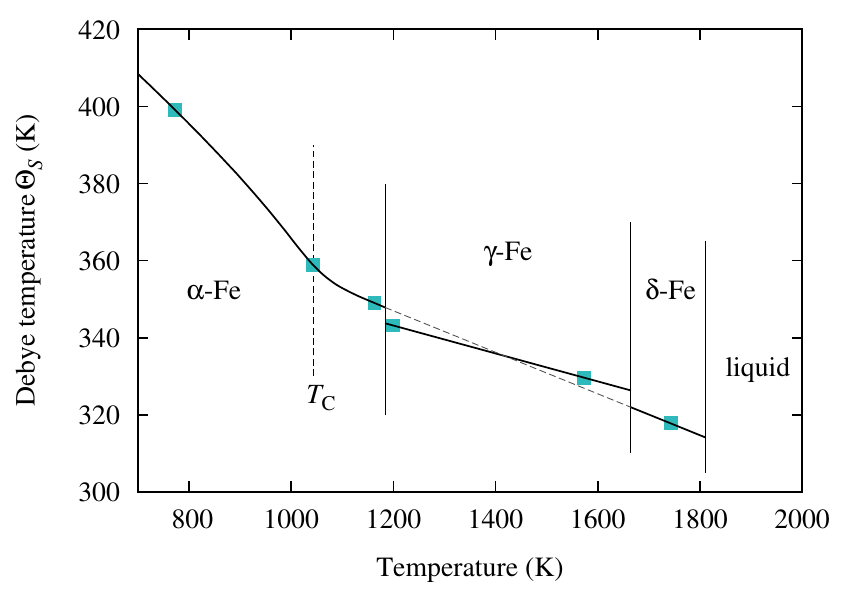}
  \caption{Debye temperature defined via the vibrational entropy as a function of temperature. The errors are smaller than the symbols.}
\label{debye_S}
\end{figure}

For settling the question for the reason of the existence of the $\delta$-phase by experiment we computed the distinct thermodynamic quantities related to the phase transitions. The most evident way to report our results is by way of Debye temperatures: We define $\Theta_U$ for a given temperature $T$ so that the internal energy of the Born-von K\'arm\'an model corresponding to the measured dispersion coincides with the internal energy of the Debye model with $\Theta_U$ as characteristic temperature. $\Theta_S$ is defined analogously via the entropy \cite{schoberlandboern1981}. The resulting values are given in Table \ref{therm}. As the temperatures of measurement are much larger than $\Theta_i$, the harmonic assumption would imply constant Debye temperatures. This is clearly not the case; the phonon softening discussed above leads to decreasing Debye temperatures with increasing temperature. Moreover, a fit with a phenomenological model for the respective structures (see the supplemental material \cite{supp} for a detailed discussion) given in figure \ref{debye_S} implies that $\Theta_S$ is discontinuous at the phase transitions, which is to be expected for a first-order transition. Our measurements constitute the first experimental determination of the Debye temperature for the high-temperature $\delta$-phase (cp.\ Ref.~\onlinecite{grimvallphysscr1976} for the previous uncertainties).

\begin{table}
  \setlength{\tabcolsep}{0pt}
    \begin{tabular}{r@{\hskip 1em}rl@{\hskip 1em}rl@{\hskip 1em}rl@{\hskip .5em}rl@{\hskip .5em}rl@{\hskip .5em}rl}
\hline\hline
    \multicolumn{1}{c}{$T$}&\multicolumn{2}{c}{$\Theta_U$}&\multicolumn{2}{c}{$\Theta_S$}&\multicolumn{2}{c}{$C_{11}$}&\multicolumn{2}{c}{$C_{44}$}&\multicolumn{2}{c}{$C'$}&\multicolumn{2}{c}{$A$}\\
    \multicolumn{1}{c}{(K)}&\multicolumn{2}{c}{(K)}&\multicolumn{2}{c}{(K)}&\multicolumn{2}{c}{(GPa)}&\multicolumn{2}{c}{(GPa)}&\multicolumn{2}{c}{(GPa)}&\\
    \hline
	773&399.0&(15)&398.5&(15)&212&(8)&112&(5)& 37&(4)&3.1&(4)\\
	1043&363.6&(15)&358.5&(15)&189&(7)&107&(4)& 16&(3)&7.0&(16)\\
	1173&354.5&(11)&348.2&(12)&190&(7)&118&(5)& 12&(3)&10.7&(33)\\
    \hline
	1200&345.3&(19)&342.8&(19)&188&(5)& 87&(3)& 16&(2)&5.7&(9)\\
	1573&333.5&(24)&329.5&(23)&171&(5)& 68&(3)& 18&(2)&3.8&(6)\\
    \hline
	1743&324.1&(7)&316.6&(6)&158&(4)& 86&(2)& 11&(1)&8.2&(6)\\
\hline\hline
    \end{tabular}
    \caption{Various properties deduced from the Born-von K\'arm\'an parameters for the respective temperatures: The Debye temperatures defined via internal energy ($\Theta_U$) and entropy ($\Theta_S$), the three cubic elastic constants and the anisotropy coefficient.}
\label{therm}
\end{table}

The recommended values for the experimental total latent heat are \quant{\Delta U_\text{tot}^{\alpha\to\gamma}=T\Delta S_\text{tot}^{\alpha\to\gamma}=0.091}{$k_\text{B}T$/atom} and \quant{\Delta U_\text{tot}^{\gamma\to\delta}=0.060}{$k_\text{B}T$/atom} \cite{chenjphaseequilbria2001}. Extrapolating our data to the transition temperatures gives the respective contributions of the vibrational entropies as \quant{\Delta S_\text{vib}^{\alpha\to\gamma} = 0.038(19)}{$k_\text{B}$/atom} and \quant{\Delta S_\text{vib}^{\gamma\to\delta} = 0.055(22)}{$k_\text{B}$/atom}. In contrast, the differences in internal vibrational energy are only \quant{\Delta U_\text{vib}^{\alpha\to\gamma} = 0.007(2)}{$k_\text{B}T$/atom} and \quant{\Delta U_\text{vib}^{\gamma\to\delta} = 0.002(2)}{$k_\text{B}T$/atom}. These figures show that at the $\alpha\to\gamma$ transition about $\Delta U_\text{elec}=\Delta U_\text{tot}-\Delta U_\text{vib}=0.08\,k_\text{B}T=9\,\text{meV}$ per atom are taken up by the electronic system, as the bcc-phase is energetically still stabilized by magnetic fluctuations. This increase in internal energy is compensated by an increase in entropy, to which the phononic subsystem contributes slightly less than half of the value, the rest being made up by electronic contributions (due to the loss of correlations). At the $\gamma\to\delta$ transition the situation is different: due to the increased temperature, the stabilizing effect of the magnetic fluctuations is lost, and the high-temperature phase again costs in internal electronic energy ($\Delta U_\text{tot}\approx\Delta U_\text{elec}=0.06k_\text{B}T$ per atom). Comparing our deduced value of \quant{\Delta S_\text{vib}=0.055}{$k_\text{B}$/atom} with the total entropy difference of \quant{\Delta S_\text{tot}=0.060}{$k_\text{B}$/atom} shows that this transition is now driven nearly exclusively by the increased vibrational entropy of the open bcc structure. The smallness of the electronic contribution to the entropy is probably due to the electronic structures of both phases being only weakly correlated. Note that our determination of $\Delta S_\text{vib}^{\gamma\to\delta}$ probably even underestimates the actual value, as in the high-temperature bcc phases typically a hardening of selected phonons with increasing temperature is found, resulting in \emph{increasing} Debye temperatures \cite{heimingprb1991}. 

\section{Conclusions}
In conclusion we find, by measuring for the first time the phonon response in the high temperature bcc phase of Fe, that the stabilization of the $\delta$-phase is due to the vibrational entropy of transverse phonons of particular low energy, favoring the picture of a first order transition driven by vibrational entropy \cite{krumhanslprb1989,kerrprb1993,mohnjphyscondmat1996}. This result is in full accordance with what we have found for the bcc phases in the nonmagnetic group 3 and 4 metals, but is more surprising for $\delta$-Fe, where magnetic fluctuations have been suspected to stabilize the body-centered cubic phase \cite{hasegawaprl1983}. Note that also the high-temperature bcc phase of Ce, another example of a system with a complex phase diagram due to magnetic interactions, has experimentally been found to be stabilized by vibrational entropy \cite{manleyphilmaglett2000,nicolauseurphysjb2001}, giving weight to the hypothesis that \emph{in general}, the existence of high-temperatur bcc phases is due to vibrational entropy. 

For the low temperature bcc structure of Fe we find that magnetic contributions establish the ferromagnetic ground state and are responsible for the structural change in the paramagnetic regime, but also for this transition there is a significant vibrational contribution to stabilize fcc-Fe.

\section*{Acknowledgments}

We would like to thank E.\ G\"unther and U.\ E\ss mann from the Max-Planck-Institut f\"ur Metallforschung, Stuttgart, for providing the $\alpha$-Fe single crystals, A.\ Krimmel for assistance during the measurements at the HMI, Berlin, and I.\ Leonov and D.\ Vollhardt for discussions and communication of theoretical results. Financial support of the LLB by the HCM program under contract No. ERB CHGECT 920001 and DFG under project PE580/3-1 and the Collaborative Research Center TRR 80 are acknowledged. 

\section*{Note added in proof}

After acceptance of this paper, a theoretical work on the basis of dynamical mean-field theory was reported \cite{leonovarxiv2014}, which confirms our interpretation.


\end{document}